\documentclass[prl,epsfig,article,superscriptaddress,tightenlines,twocolumn]{revtex4}
\usepackage{latexsym,amsmath,tabularx,amssymb,bm}
\usepackage{graphicx}

\usepackage{psfrag}
\usepackage{color}

\newcommand \bea{\begin{eqnarray}}
\newcommand \eea{\end{eqnarray}}

\def\simge{\mathrel{%
   \rlap{\raise 0.511ex \hbox{$>$}}{\lower 0.511ex \hbox{$\sim$}}}}
\def\simle{\mathrel{
   \rlap{\raise 0.511ex \hbox{$<$}}{\lower 0.511ex \hbox{$\sim$}}}}

\def\simle{\mathrel{
   \rlap{\raise 0.511ex \hbox{$<$}}{\lower 0.511ex \hbox{$\sim$}}}}

\def\simge{\mathrel{%
    \rlap{\raise 0.511ex \hbox{$>$}}{\lower 0.511ex \hbox{$\sim$}}}}
\def\simle{\mathrel{
    \rlap{\raise 0.511ex \hbox{$<$}}{\lower 0.511ex \hbox{$\sim$}}}}
\newcommand{\beq}{\begin{eqnarray}}
\newcommand{\eeq}{\end{eqnarray}}

\usepackage{xcolor}
\begin{document}

\title{Finite scale singularity in the renormalization group flow of a reaction-diffusion system}

\author{Damien Gredat}
\affiliation{Service de Physique de l'Etat Condens\'e, CEA Saclay
CNRS URA~2464, 91191 Gif-sur-Yvette, France}
\author{Hugues Chat\'e}
\affiliation{Service de Physique de l'Etat Condens\'e, CEA Saclay
CNRS URA~2464, 91191 Gif-sur-Yvette, France}
\affiliation{Laboratoire de Physique Th\'eorique de la Mati\`ere Condens\'ee, UPMC, CNRS UMR 7600, 4 Place Jussieu, 75252 Paris, France}
\author{Ivan Dornic}
\affiliation{Service de Physique de l'Etat Condens\'e, CEA Saclay
CNRS URA~2464, 91191 Gif-sur-Yvette, France}
\author{Bertrand Delamotte}
\affiliation{Laboratoire de Physique Th\'eorique de la Mati\`ere Condens\'ee, UPMC, CNRS UMR 7600, 4 Place Jussieu, 75252 Paris, France}

%%%%%%%%%%%%%%%%%%%%%%%%%%%%%%%%%%%%%%%%%%%%%%%%%%%%%%%%%%%%%%%%%%%%%%%%%%%%%%%%%%%%%%%%%%%%%%%%%%%%%%%%%%%%%%%%%%%%%%%%%%%%
\begin{abstract}
We study the nonequilibrium critical behavior of the pair contact process with diffusion (PCPD) by means of  
nonperturbative functional renormalization group techniques. We show that usual perturbation theory fails because 
the effective potential develops a nonanalyticity at a finite length scale:
%We also demonstrate that 
Perturbatively forbidden terms are dynamically
generated and the flow can be continued  once they are taken into account.
Our results suggest that the critical behavior of PCPD can be either in the directed percolation 
or in a new (conjugated) universality class.

\end{abstract}
%%%%%%%%%%%%%%%%%%%%%%%%%%%%%%%%%%%%%%%%%%%%%%%%%%%%%%%%%%%%%%%%%%%%%%%%%%%%%%%%%%%%%%%%%%%%%%%%%%%%%%%%%%%%%%%%%%%%%%%%%%%%

 % end of PACS codes

\date{\today}% It is always \today, today,
               %  but any date may be explicitly specified

\maketitle
%%%%%%%%%%%%%%%%%%%%%%%%%%%%%%%%%%%%%%%%%%%%%%%%%%%%%%%%%%%%%%%%%%%%%%%%%%%%%%%%%%%%%%%%%%%%%%%%%%%%%%%%%%%%%%%%%%%%%%%%%%%%%%%%%%%%%%%%%%%%%%

Reaction-diffusion systems involving one particle species (also known as branching and annihilating random walks (BARW))
are stochastic out of equilibrium systems important from both a phenomenological and a theoretical viewpoint.
They consist of identical particles $A$, diffusing on a $d$-dimensional lattice, that can branch ($n A\to (n+p)A$) or annihilate
($nA\to (n-q)A$). The competition 
between these two types of reaction is generically
responsible for the existence of transitions between an active 
phase where the density of particles is finite,
and an ``absorbing" phase
where all particles, and thus all fluctuations, have disappeared.
Such models provide the building  blocks of a large variety of applications and models in physics and beyond, 
and are therefore of fundamental importance
% to understand the mechanisms at stake in the simplest possible situations 
\cite{HinOdo_abs}.
They also have the advantage of providing a relatively simple theoretical framework 
for the study of the different universality classes of absorbing phase transitions.

Our understanding of out-of-equilibrium critical phenomena in general and absorbing phase transitions
in particular has benefited from perturbative approaches  \cite{CarTau98,Janssen81}, 
%over the last few years, 
but important advances were brought recently by the application of 
nonperturbative renormalization group (NPRG) methods \cite{Can04,Can05,Can10, Can05r}. 
For the two prominent cases of BARW,
%have been studied both by perturbative \cite{CarTau98} and nonperturbative RG methods \cite{Can05r}: 
$A\to 2A$, $2A \to \emptyset$
which represents the directed percolation (DP) class, and 
$A\to 3A$, $2A \to \emptyset$ which belongs to the parity-conserving, 
or generalized voter class, the success of the NPRG owed to
the presence of nonperturbative features.

The case of the ``pair contact process with diffusion'' (PCPD) has largely resisted analysis so far \cite{HenHin04}
but was not studied with NPRG methods.
The simplest BARW model in this class consists 
%of the set
of reactions $2A\to3A$ and $2A \to \emptyset$ with rates $\sigma$ and $\lambda$. 
(A limiting reaction such as $3A \to \emptyset$, with rate $\lambda'$, is actually needed  to ensure a finite density active
phase \footnote{When this reaction is absent
the model is soluble but this solution does not help understanding the phase transition we study here, see  
M.J. Howard and U.C. T\"auber, J. Phys. A: Math. Gen. {\bf 30}, 7721 (1997);
M. Paessens and G. M. Sch\"utz, J. Phys. A: Math. Gen. {\bf 37}, 4709 (2004).}.)
The distinctive feature of the PCPD is that two particles must meet 
to trigger branching. 
On general grounds, this is not expected to be a relevant ingredient defining 
universality classes, hence the interest raised by results obtained so far on the 
critical behavior of PCPD: It has been intensively studied numerically in $d=1$ but 
remains unclear because of the presence of slow dynamics and/or strong corrections to scaling \cite{ScalCorr_PCPD}. 
The debate, ongoing still recently, is to know whether PCPD belongs to the DP universality class 
\cite{BarCar03,ScalCorr_PCPD,BarSch12} or not \cite{PCPD_NClass,Par05,Par09,Par12}. 
%For instance, in \cite{BarSch12}
%the critical exponents found are claimed to be consistent with those of DP whereas in \cite{Par05,Par12} 
%they are found different and interpreted as a signal of a new universality class. 

Even the status of $d_{\rm c}$, the upper critical dimension of PCPD, is unclear: numerically, 
$d=3$ seems beyond it \cite{DCM05}, but in $d=2$,
the presence of large corrections to scaling is difficult to disentangle from logarithmic terms preventing clear conclusions
in spite of indications of mean-field behavior \cite{Odo04,Odo02}. 
In perturbation theory the RG flow of PCPD
goes to the Gaussian fixed point for $d>2$ and sufficiently small coupling constants, whereas it
blows up at a finite scale for larger couplings or for $d<2$ \cite{Wij04}. This suggests $d_{\rm c}=2$. 
Note that the explosive flow forbids the exploration of the long-distance physics of the model. 
This is also known to occur in quantum chromodynamics 
(at the confinement scale), in the $O(N)$ nonlinear sigma model (at the scale of the correlation length) \cite{polyakov}, 
and in pinned elastic manifolds (at the Larkin length) \cite{Cusp_RM1}. 
In this last case, NPRG methods at the functional level allowed to treat the problem  \cite{Cusp_RM2}.
%In the latter case, the flow can be continued if one works 
%with the functional RG \cite{Cusp_RM}. 

In this Letter, we examine the PCPD field theory in the light of the NPRG, 
explain why perturbation theory fails, and how to avoid its problems. We 
show that the potential 
in the running effective action develops a singularity at a finite scale, 
signalling that couplings that are perturbatively forbidden
are dynamically generated. Once taken into account
 the RG flow can be continued and a fixed point can be found. 
Our results suggest that the critical behavior
of the model is either in the DP class or possibly in a new class
characterized by a ``conjugated" symmetry,
and that $d_{\rm c}=2$ only at small coupling.
%More generally, o
Our study indicates that NPRG is a powerful tool for dealing with
similar situations beyond reaction-diffusion systems.

%%%%%%%%%%%%%%%%%%%%%%%%%%%%%%%%%%%%%%%%%%%%%%%%%%%%%%%%%%%%%%%%%%%%%%%%%%%%%%%%%%%%%%%%%%%%%%%%%
\textit{ The field theory associated with PCPD}. By using the usual Doi-Peliti formalism \cite{Doi76}
it is possible to derive the action 
associated with PCPD from first principles:
%%%%%%%%%%%%%%%%%%%%%%%%%%%%%%%%%%%%%%%%%%%%%%%%%%%%%%%%%%%%%%%%%%%%%BE
\begin{eqnarray}
 {\cal S}=\int_{\bf x}&&\!\!\!\left[\bar\phi\left(\partial_t \phi - D \nabla^2\phi\right) + 
\phi^2\left(g_1\bar\phi+ g_2 \bar\phi^2+ g_3 \bar\phi^3\right)\right.\nonumber \\
&&\left. +\phi^3\left(3\lambda' \bar\phi + O(\bar\phi^2)\right)\right]
\label{action}
\end{eqnarray}
%%%%%%%%%%%%%%%%%%%%%%%%%%%%%%%%%%%%%%%%%%%%%%%%%%%%%%%%%%%%%%%%%%%%%EE
where $\phi$ and $\bar\phi$ are complex conjugates, ${\bf~x}=(t,\vec{x})$, ${\int}_{\bf x}={\int}d^dx~dt$, and 
$g_1=2\lambda-\sigma$, $g_2=\lambda-2\sigma$, $g_3=-\sigma$. 
%Two types of 
%perturbative calculations have been performed from this field theory in \cite{Wij04} 
%depending on the dimensions given to the fields  $\phi$ and $\bar\phi$. 
%In the first calculation where $[\phi]=\kappa^d$ and $[\bar\phi]=\kappa^0$ where
%$\kappa$ is a momentum scale, all monomials $\phi^2(\bar\phi)^n$ with $n\ge 1$ are equally relevant and the flows
%of all the corresponding couplings have a priori to be considered. In the other calculation where $[\phi]=\kappa^{d/2}=[\bar\phi]$
%only the terms $\phi^2\bar\phi$ and $\phi^2\bar\phi^2$ are relevant and need to be considered perturbatively (together
%with the term $\phi^3\bar\phi$). 
Within perturbation theory, one finds that initializing the RG flow with  the (bare) couplings 
of the action ${\cal S}$, Eq.(\ref{action}), the running coupling $g_2(k)$ ($k$ being a momentum scale) 
diverges at a finite scale $k_{\rm c}$ for $d<2$ which bars
from exploring scales below $k_{\rm c}$. Moreover, the $\langle\bar\phi\phi \rangle$ response function 
does not receive any loop correction and thus the dynamical exponent $z$ remains equal to 2
which is clearly invalidated by numerical results. For $d>2$,
all couplings are irrelevant around the Gaussian fixed point which is thus (locally) attractive. 
(Hence the conclusion that $d_{\rm c}=2$.) 
%is therefore interpreted as the upper critical dimension (however, see below). 
We now present a calculation where  the flow no longer diverges but
develops a singularity that can naturally be taken into account at the price of
working functionally.

%%%%%%%%%%%%%%%%%%%%%%%%%%%%%%%%%%%%%%%%%%%%%%%%%%%%%%%%%%%%%%%%%%%%%%%%%%%%%%%%%%%%%%%%%%%%%%%%%%%%%%%%%%%%%%%%%%%%%%%%%%%%%%%%%%%%%%%%%%%%%%
\textit{ The nonperturbative renormalization group.} NPRG follows Wilson's idea \cite{WilKog74} of partial integration over fluctuations.
It builds a one-parameter family of models indexed by a momentum scale $k$ such  that fluctuations are 
smoothly included as $k$ is lowered from the inverse lattice spacing $\Lambda$ down to $k=0$ where they have all 
been summed over. To this aim, we add to the original action a momentum-dependent mass-like term \cite{Ber02,Del07}
%%%%%%%%%%%%%%%%%%%%%%%%%%%%%%%%%%%%%%%%%%%%%%%%%%%%%%%%%%%%%%%%%%%%%BE
\begin{equation}
 \Delta  {\cal S}_k=\frac{1}{2}\int_{\bf q}\phi_i({\bf -q})\, [R_k(q)]_{ij}\, \phi_j({\bf q})
\end{equation}
%%%%%%%%%%%%%%%%%%%%%%%%%%%%%%%%%%%%%%%%%%%%%%%%%%%%%%%%%%%%%%%%%%%%%EE
where ${\bf q}=(\omega,\vec{q})$, $q=\vert\vec{q}\vert$,
 $i=1,2$, $\phi_1=\phi$, $\phi_2=\bar\phi$ and repeated indices are summed over. With, {\it e.g.},  
$[R_k]_{12}=[R_k]_{21}=(k^2-q^2)\theta(k^2-q^2)$ and $[R_k]_{11}=[R_k]_{22}=0$,
the fluctuation modes $\phi_i(q>k)$ 
are unaffected by $\Delta {\cal S}_k$ while the others with $q<k$ are essentially frozen. The $k$-dependent generating functional
of correlation and response functions thus reads
%%%%%%%%%%%%%%%%%%%%%%%%%%%%%%%%%%%%%%%%%%%%%%%%%%%%%%%%%%%%%%%%%%%%%BE
\begin{equation}
 {\cal Z}_k[J_1,J_2]=\int D\phi D\bar\phi\, e^{-S-\Delta S_k+\int_{\bf x}J_i\phi_i} .
\end{equation}
%%%%%%%%%%%%%%%%%%%%%%%%%%%%%%%%%%%%%%%%%%%%%%%%%%%%%%%%%%%%%%%%%%%%%EE
The effective action $\Gamma_k[\psi_i]$, where $\psi_1=\psi=\langle\phi_1\rangle$, 
$\psi_2=\bar\psi=\langle\phi_2\rangle$,
  is given by the Legendre transform of 
${\cal W}_k=\log{\cal Z}_k$ (up to the term proportional to $R_k$):
%%%%%%%%%%%%%%%%%%%%%%%%%%%%%%%%%%%%%%%%%%%%%%%%%%%%%%%%%%%%%%%%%%%%%BE
\begin{equation}
 \Gamma_k[\psi,\bar\psi]+ {\cal W}_k=\int_{\bf x}J_i\psi_i -\int_{\bf q} R_k(q) \psi({\bf q})\bar\psi({\bf -q}).
\end{equation}
%%%%%%%%%%%%%%%%%%%%%%%%%%%%%%%%%%%%%%%%%%%%%%%%%%%%%%%%%%%%%%%%%%%%%EE
The two-point functions can be computed from  $\Gamma_k$ by differentiating:
%%%%%%%%%%%%%%%%%%%%%%%%%%%%%%%%%%%%%%%%%%%%%%%%%%%%%%%%%%%%%%%%%%%%%BE
\begin{equation}
[\,\Gamma_k^{(2)}\,]_{i_1i_2}[{\bf x}_1,{\bf x}_2,\psi,\bar\psi]=
\frac{\delta^{2}\Gamma_k}{\delta\psi_{i_1}({\bf x}_1)\delta\psi_{i_2}({\bf x}_2)}
\end{equation}
%%%%%%%%%%%%%%%%%%%%%%%%%%%%%%%%%%%%%%%%%%%%%%%%%%%%%%%%%%%%%%%%%%%%%EE
and the exact flow equation for $\Gamma_k[\psi,\bar\psi]$ reads \cite{Ber02}:
%%%%%%%%%%%%%%%%%%%%%%%%%%%%%%%%%%%%%%%%%%%%%%%%%%%%%%%%%%%%%%%%%%%%%BE
\begin{equation}
\label{eq-wetterich}
\partial_k \Gamma_k= \frac{1}{2}{\rm Tr}\int_{\bf q} \partial_k R_k\ . \   G_k \ {\rm with}\ G_k=[\Gamma_k^{(2)}+R_k]^{-1}.
\end{equation}
%%%%%%%%%%%%%%%%%%%%%%%%%%%%%%%%%%%%%%%%%%%%%%%%%%%%%%%%%%%%%%%%%%%%%EE
When $k$ decreases from $\Lambda$ to 0, $\Gamma_k$ varies between the (bare) action: 
$\Gamma_{k=\Lambda}= {\cal S}$, and the full
effective action: $\Gamma_{k=0}=\Gamma$. Solving the flow equation (\ref{eq-wetterich}) 
is thus equivalent to solving the model. This is however impossible to do exactly 
and approximations must be made. We perform here
the local potential approximation (LPA) which is known to work well for 
the determination of the critical behavior of models either 
at equilibrium or out-of-equilibrium \cite{Ber02,LPA,Can04,Can05}:
% \textcolor{red}{ and that consists 
%in focusing on the long-time and large distance behavior of correlation and response functions}. It reads:

%%%%%%%%%%%%%%%%%%%%%%%%%%%%%%%%%%%%%%%%%%%%%%%%%%%%%%%%%%%%%%%%%%%%%BE
\begin{equation}
\label{LPA}
\Gamma_k\to \Gamma_k^{\rm LPA}=\int_{\bf x}\left( \bar\psi(\partial_t - D \nabla^2)\psi + U_k(\psi,\bar\psi)\right).
\end{equation}
%%%%%%%%%%%%%%%%%%%%%%%%%%%%%%%%%%%%%%%%%%%%%%%%%%%%%%%%%%%%%%%%%%%%%EE
Substituting Eq.(\ref{LPA}) into Eq.(\ref{eq-wetterich}), choosing the $R_k$ function described
above and integrating over ${\bf q}$, we find :
%%%%%%%%%%%%%%%%%%%%%%%%%%%%%%%%%%%%%%%%%%%%%%%%%%%%%%%%%%%%%%%%%%%%%BE
\begin{equation}
\label{flot-LPA}
\!\!\partial_k  U_k=C_d\left\lgroup
\frac{{k^2}+U_k^{(1,1)}}{\sqrt{\left({k^2}+U_k^{(1,1)}\right)^2-U_k^{(0,2)}U_k^{(2,0)}}}-1\right\rgroup
\end{equation}
%%%%%%%%%%%%%%%%%%%%%%%%%%%%%%%%%%%%%%%%%%%%%%%%%%%%%%%%%%%%%%%%%%%%%EE
where $C_d=4k^{2+d}[2^{d+1}\pi^{d/2}d \Gamma(d/2)]^{-1}$ and the upper indices code for derivatives in $\psi$ and $\bar\psi$.
It is easy to verify that performing  a field expansion of $U_k(\psi,\bar\psi)$ around $(0,0)$ leads, for the couplings
in front of the monomials  $\psi^m\bar\psi^n$, to the same difficulties as encountered perturbatively: 
the flow blows up at a finite scale for $d<2$. We thus need to work functionally. 
However, we can further simplify our ansatz Eq.(\ref{LPA}) 
by performing an expansion of $U_k(\psi,\bar\psi)$
in  $\bar\psi$  only while remaining functional in the $\psi$-direction. We thus replace $U_k(\psi,\bar\psi)$ by
%%%%%%%%%%%%%%%%%%%%%%%%%%%%%%%%%%%%%%%%%%%%%%%%%%%%%%%%%%%%%%%%%%%%%BE
\begin{equation}
\label{semi-expansion-1}
 U_k(\psi,\bar\psi)\to \sum_{n=1}^{N} \bar\psi^n V_{n,k}(\psi)
\end{equation}
%%%%%%%%%%%%%%%%%%%%%%%%%%%%%%%%%%%%%%%%%%%%%%%%%%%%%%%%%%%%%%%%%%%%%EE
and we typically truncate the sum at order $N=3$ or 4. Our approximation scheme is therefore based on the assumption
that the LPA is sufficient, and that the nontrivial 
features of the model lie in the $\psi$ direction (which is not spoiled by the expansion in  $\bar\psi$).
Inserting Eq.(\ref{semi-expansion-1}) into Eq.(\ref{flot-LPA}) we find the flow equations for the 
functions $V_{n,k}$ at order $N=3$ :
%%%%%%%%%%%%%%%%%%%%%%%%%%%%%%%%%%%%%%%%%%%%%%%%%%%%%%%%%%%%%%%%%%%%%BE
\begin{align} 
\label{flot-V} 
\dot{V}_1&= B V_2 V_1'' \\
\dot{V}_2&=B\!\left[ \frac{1}{2} V_1'' (A V_2 (3 A V_2 V_1'' \!-\!8 V_2') \!+\! 6
   V_3)+V_2 V_2''\right] \\
\dot{V}_3&= B \! \left[\frac{1}{2} \left(A \left(8 V_2' \left(3 V_1'' \left(A V_2 V_2'-V_3\right)-V_2
   V_2''\right)\right.\right.\right.\nonumber \\
          &\left.\left.\left.+ V_2 V_1'' \left(A \left(V_1'' \left(A V_2 \left(5 A V_2
   V_1''-24 V_2'\right)+18 V_3\right)\right.\right.\right.\right.\right. \nonumber \\
          &\left.\left.\left.\left.+\left.6 V_2 V_2''\right)-12
   V_3'\right)\right)+6 V_3 V_2''\right)+V_2 V_3'' \phantom{\frac{1}{2}} \hspace{-.7em} \right] 
\end{align}
%%%%%%%%%%%%%%%%%%%%%%%%%%%%%%%%%%%%%%%%%%%%%%%%%%%%%%%%%%%%%%%%%%%%%EE
where, for 
%notational 
simplicity, we have omitted the index $k$ in the functions $V_{n,k}$, $V_n'$ 
and $V_n''$ are the derivatives of these functions with respect to 
%$\psi$, $\dot{V}_n= k\partial_k V_n$, $A=(1 +V_1')^{-1}$ and $B=A^{3}\vert A^{-1}\vert$.
$\psi$, $\dot{V}_n= k\partial_k V_n$, $A=( k^2 +V_1')^{-1}$ and $B=(k^{d+2}4 v_d/d)A^{3}\vert A^{-1}\vert$.
The initial conditions of these flow equations are provided by the action ${\cal S}$ with $\phi_i\to\psi_i$.
%\textcolor{red}{ Most crucially}, t
They impose that $\forall n$: $V_{n,k=\Lambda}'(\psi=0)=0$:
%which means that 
the bare potential has no linear term in $\phi$. 
Perturbatively, it is easy to show
that these terms cannot be generated in $U_k$
since all Feynman diagrams involve at least two incoming particles and thus two fields $\psi$. 

\begin{figure}[t]
\includegraphics[width=\columnwidth,clip]{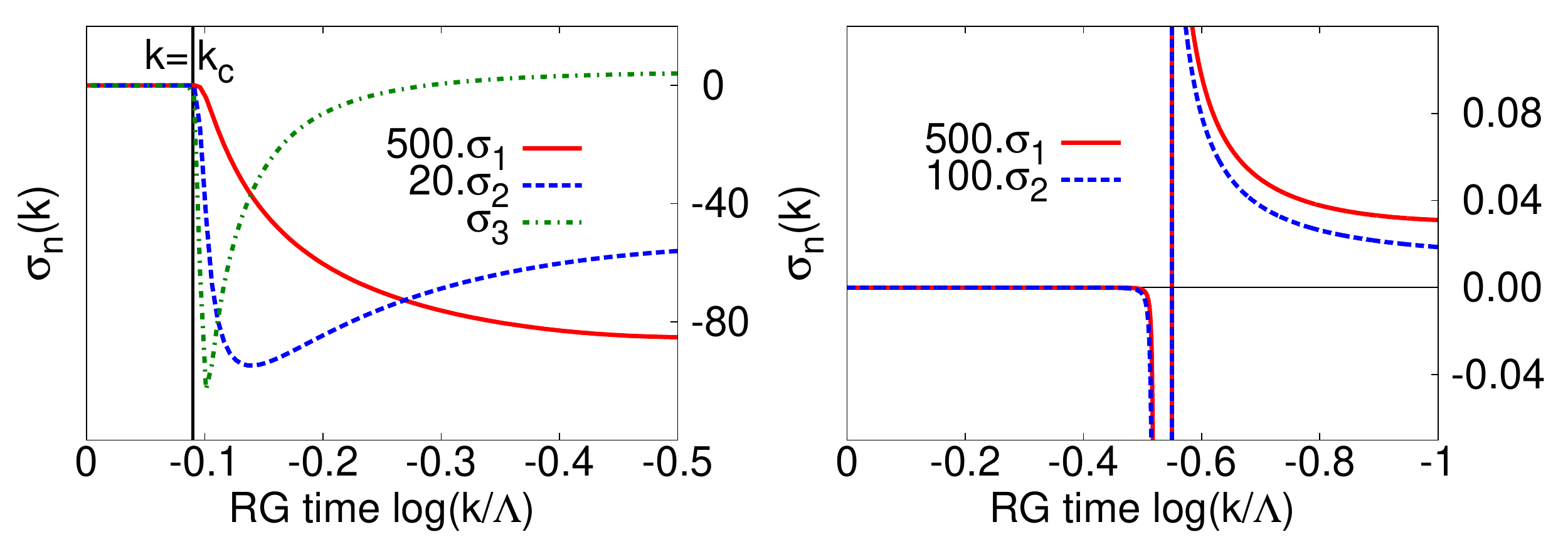} 
\caption{(color online) Left: Flows in $d=1$ of the linear couplings $\sigma_{n}(k)=V_{n,k}'(0)$ 
in the semi-functional approach, Eq.(\ref{semi-expansion-1}). 
Right: Same when  $U_k$ is expanded in both $\psi$
and $\bar\psi$ and infinitesimal linear couplings are considered
($\sigma_{1}(k=\Lambda)=0$, $\sigma_{2}(k=\Lambda)= 10^{-8}$, all other rates are of order 1).  
Only terms at most quartic in the fields have been retained.
The RG flow of all couplings remains finite but shows an abrupt change around $k_{\rm c}\simeq0.6$.}
\label{fig: sigm}
\end{figure}

We have numerically integrated the coupled flow equations 
of the functions $V_{n,k}(\psi)$ with $N= 4$ (see Eqs.(\ref{flot-V}) for the $N=3$ case) 
together with the initial conditions provided by ${\cal S}$, Eq.(\ref{action}).
In the early stage of this flow, 
%that is for $k$ a little lower than $\Lambda$, 
the linear term of each of these functions remains identically zero as
naively expected. However, at a finite scale $k_{\rm c}$ which is typically the scale where the 
perturbative flow blows up, 
a linear term is generated in all these functions (Fig.\ref{fig: sigm}): the potential $U_k$ develops a {\it corner} at 
$\psi=0$ and its analytic structure is changed below $k_{\rm c}$.

%%%%%%%%%%%%%%%%%%%%%%%%%%%%%%%%%%%%%%%%%%%%%%%%%%%%%%%%%%%%%%%%%%%%%BF
%%%%%%%%%%%%%%%%%%%%%%%%%%%%%%%%%%%%%%%%%%%%%%%%%%%%%%%%%%%%%%%%%%%%%EF

\begin{figure}[t]
\includegraphics[width=0.9\columnwidth,clip]{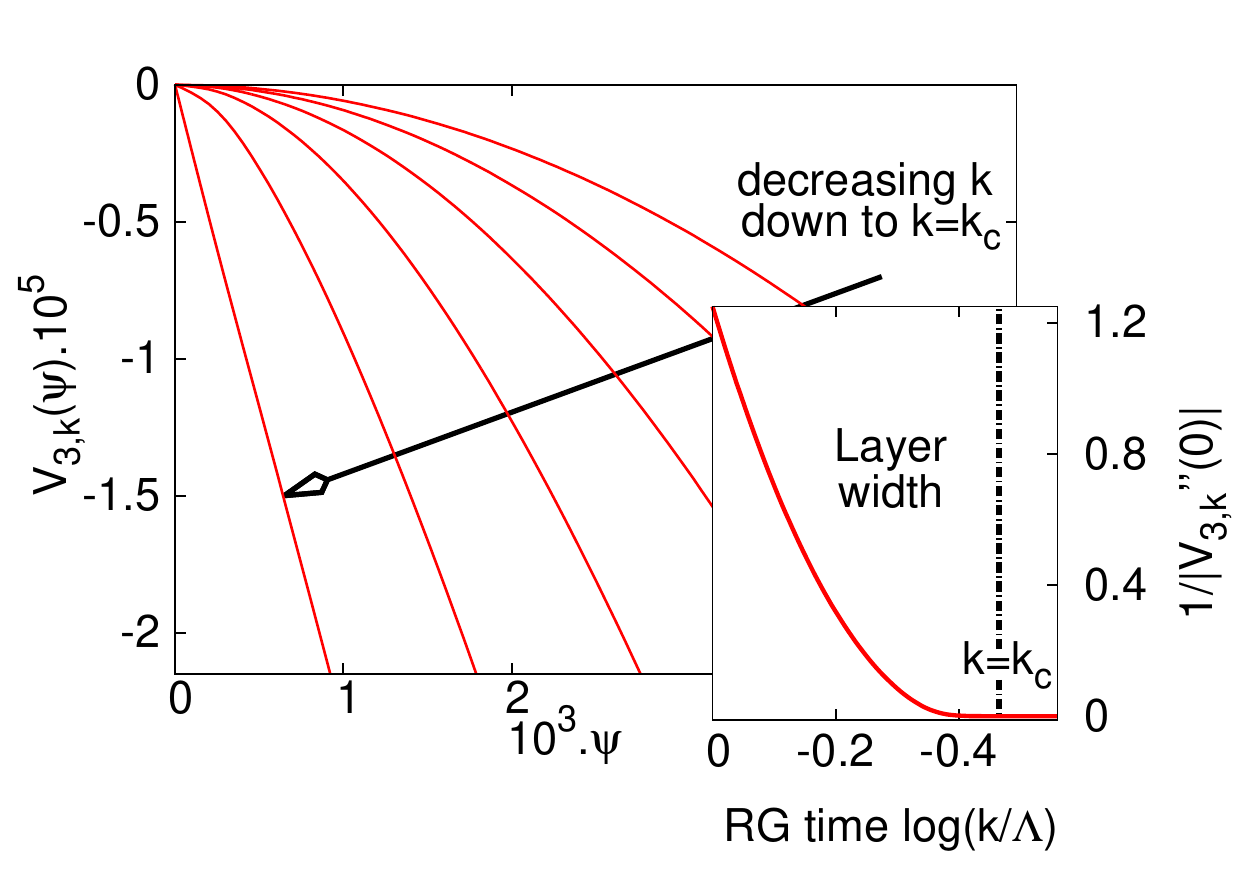}
\caption{(color online)
Boundary layer for $V_{3,k}(\psi)$, whose behavior around the origin is displayed
for $k\simeq k_{\rm c}$ and several values of $k$ above $k_{\rm c}$. 
For all $k>k_{\rm c}$, $V_{3,k}$ starts quadratically at $\psi=0$ whereas a 
linear term appears at $k=k_{\rm c}$. Inset: width of the boundary layer as a function
of $\log k/\Lambda$.}
\label{fig: v1}
\end{figure}

A detailed study of the emergence of the linear terms reveals that for $k\gtrsim k_{\rm c}$, 
a boundary layer appears in the $V_{n,k}(\psi)$  functions  such that in the inner part of the layer, 
that is, at small $\psi$, these functions are expandable around $\psi=0$ (for $\psi\ge 0$)
and   start quadratically in  $\psi$. In the outer region, a linear part appears in the $V_{n,k}(\psi)$ functions. 
As $k$ approaches $k_{\rm c}$ from above, the width of the layer
decreases and vanishes at $k_{\rm c}$, leaving the linear term as the dominant term around $\psi=0$ 
(Fig.\ref{fig: v1}). Below $k_{\rm c}$, the linear terms remain present in the vicinity of $\psi=0$ and 
the flow can be continued all the way to $k=0$. 
%We have verified that for generic initial 
%conditions, this behavior exists in all dimensions (see however below)
%and is stable when we truncate the $\bar\psi$-expansion of $U_k$ either at order $N=3$ or 4, the difference
%being only quantitative. 
We have checked that this emerging scenario holds at order $N=3$ and 4 of the 
$\bar\psi$-expansion. Converged, higher-order results are unfortunately
difficult to obtain \footnote{They
rely on a rather delicate numerical analysis because 
it is not easy to characterize reliably the emergence of nonanalytic behavior in $\psi$ at a finite scale 
using  grids both in the RG ``time" $\log k/\Lambda$ and in $\psi$.}.

%%%%%%%%%%%%%%%%%%%%%%%%%%%%%%%%%%%%%%%%%%%%%%%%%%%%%%%%%%%%%%%%%%%%%BF
%%%%%%%%%%%%%%%%%%%%%%%%%%%%%%%%%%%%%%%%%%%%%%%%%%%%%%%%%%%%%%%%%%%%%EF

We have confirmed our scenario by considering
% PCPD as the limit of 
the usual set of reactions $2A\to3A$, $2A \to \emptyset$ and $3A \to \emptyset$ 
complemented by $ A\to 2A$ and $ A\to \emptyset$ with infinitesimal rates.
Expanding the potential $U_k$ in both $\psi$ and $\bar\psi$ we now have (at least) 
two new couplings that are linear in $\psi$: 
$\sigma_1(k) \psi\bar\psi$ and $\sigma_2(k) \psi\bar\psi^2$
with $\sigma_{1,2}(k=\Lambda)$ infinitesimal. If the limit $\sigma_{1,2}(k=\Lambda)\to 0$ were regular
 then $\sigma_{1,2}(k)$ would start playing a significant role in the RG flows of the other couplings
only below a very small scale $k$  that would
go to 0 as $\sigma_{1,2}(k=\Lambda)\to 0$.  We find on the contrary 
that while for $k$ just below $\Lambda$ the flows of all the couplings are indeed almost insensitive to 
$\sigma_{1,2}$  (when they are initially extremely small), this is no longer the case 
for $k\simeq k_{\rm c}$  since the dramatic increase of $g_2(k)$ 
makes $\sigma_{1,2}(k)$ grow abruptly around $k_{\rm c}$, independently of their initial smallness, see Fig. \ref{fig: sigm}.
%This phenomenon occurs independently of their initial values because 
%the smaller they are initially the larger is the increase of $g_2(k\simeq k_{\rm c})$. 
For $k$ close to $k_{\rm c}$, the back reactions of $\sigma_{1,2}(k)$ on the flows of 
all the  other couplings start to be 
significant and eventually modify them completely  since  $\sigma_{1,2}$ are the most relevant couplings.
The RG flow is no longer singular ($g_2(k_{\rm c})$ remains finite) but lives  below $k_{\rm c}$ in a larger functional space
involving the couplings linear in $\psi$. This result is fully consistent 
with what is found in the functional viewpoint.

\textit{Criticality in $d=1$}. Our two different approaches both conclude that terms linear in $\psi$ are generated below 
a nonuniversal scale $k_{\rm c}$ in $d=1$. We can therefore consider the field theory  obtained just below $k_{\rm c}$ as a new field theory 
that can be studied per se, the difficulty being that its action is non polynomial since the functions $V_{n,k}(\psi)$ are not. 
In a perturbative analysis only the terms of lowest degrees in $\psi$ and $\bar\psi$ would be retained 
in the bare action, {\it i.e.}
$\psi\bar\psi$, $\psi^2\bar\psi$ and $\psi\bar\psi^2$. Depending on the relative sign of the 
two cubic terms, this action, truncated at order three,
%in the fields, 
exhibits one of the following symmetries (after a trivial rescaling of the fields): 
$\bar{\psi}(t)\rightleftarrows \pm\psi(- t)$. The 
minus sign corresponds to the cubic terms having opposite signs. This is the ``rapidity" symmetry defining the DP class.
The other sign defines a new, ``conjugated'', symmetry. We call the corresponding class DP'.
It is easy to show in our framework, Eq.(\ref{flot-LPA}), that when only the above terms are kept in $U_k$,
only two nontrivial fixed points exist and that they show either one or the other of the above symmetries
(we call them DP and DP').
No such result exists beyond this simple truncation. To the best of our knowledge, the DP' symmetry has never appeared 
nor been studied before.
%in particular because a general potential does not exhibit any symmetry. 
%We now show that at criticality the PCPD flow is indeed attracted towards the DP fixed point.

%%%%%%%%%%%%%%%%%%%%%%%%%%%%%%%%%%%%%%%%%%%%%%%%%%%%%%%%%%%%%%%%%%%%%BF
\begin{figure}[t]
 \includegraphics[width=\columnwidth,clip]{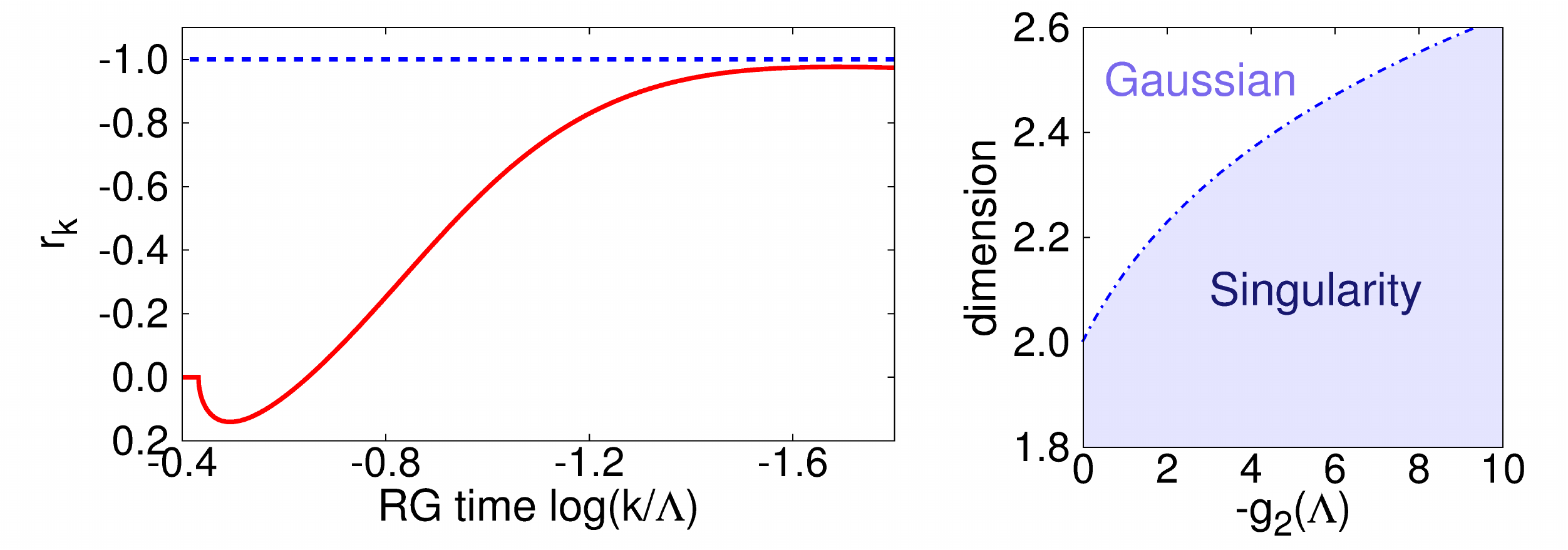}
\caption{(color online) Left: Flow of the symmetry-testing ratio $r_k$ computed in the semi-functional approach with $N=4$. For
$k<k_{\rm c}$, $r_k$ flows toward the DP value $r_k^{(DP)}=-1$. Right: line separating the basin of attraction 
of the Gaussian fixed point and the domain where the singularity is present in the $(d,g_2(\Lambda))$ plane when $g_1(\Lambda)=0$.}
\label{fig: DPcond}
\end{figure}
%%%%%%%%%%%%%%%%%%%%%%%%%%%%%%%%%%%%%%%%%%%%%%%%%%%%%%%%%%%%%%%%%%%%%EF

Two difficulties appear when studying the existence and the nature of the fixed point. 
First, when using a complete field expansion of $U_k$, it is difficult to initialize the flow below $k_{\rm c}$ 
with what has been found functionally using Eq.(\ref{semi-expansion-1})
just after the singularity because this amounts to projecting functions of
$\psi$ onto polynomials, setting infinitely many terms of high degree to 0. Second, if we work directly with our semi-functional 
approximation, Eq.(\ref{semi-expansion-1}), 
the roles of $\psi$ and $\bar\psi$ are dissymmetric which hinders the search of a fixed point 
exhibiting a symmetry that exchanges  $\psi$ and $\bar{\psi}$ as in DP or DP'. 
We have nevertheless studied the existence of a fixed point using both approaches. 
Within richer and richer polynomial approximations of $U_k$ (where $\psi$ and $\bar\psi$ play
symmetric roles) initialized with 
couplings found with the semi-functional approximation just after the singularity, we have found a range of initial
reaction rates leading to the DP fixed point. In the other, semi-functional approximation truncated at order $N=4$,
 we have recorded the flow of
$r_k= (U_k^{(2,1)}/U_k^{(1,2)})\sqrt{U_k^{(1,3)}/U_k^{(3,1)}}$ (calculated at $\psi=\bar\psi=0$).
This quantity is an indicator of the nature of 
the fixed point because if $r_k=-1$ the
expansion of $U_k$  up to order four in the fields exhibits the rapidity symmetry 
(up to a rescaling of the fields). If $r_k=+1$ the same holds true for the DP' symmetry. Notice that
just below $k_{\rm c}$, $r_k$ is neither 1 nor $-1$.
We show in Fig.\ref{fig: DPcond} the evolution of $r_k$ computed from  Eq.(\ref{semi-expansion-1}) with $N=4$. 
Although the  instabilities of our numerical code prevent us from finding a true fixed point value for $r_k$ there is little doubt
that it indeed reaches a plateau at the value $-1$ so that the expected fixed point should be that of DP in agreement with what is found
in the polynomial approximation. Notice however that if we work with a field truncation of $U_k$, starting at $k=\Lambda$
with a very small coupling $\sigma_{2}(k=\Lambda)$ as explained above, we find, at least in the simplest truncation of $U_k$,
either the DP or DP' fixed points
depending on the initial rates (we have not been able to confirm this result with truncations involving 
higher powers of the fields because of the extreme sensitivity of the flow at small $k$ on the choice of the
initial rates). Interestingly, we find that the vicinity of both fixed points is reached after a very long
RG ``time" ($\log k/\Lambda\sim -9$) which means that the scaling regime only appears at very large lengthscales. 
This could explain why it is so difficult to observe the asymptotic regime in numerical simulations. 
Our general conclusion is therefore that the critical behavior of PCPD in $d=1$ 
should be either in the DP or DP' universality class and our best results are in favor of DP.

\noindent\textit{The upper critical dimension.}
The  determination of $d_{\rm c}$ is the same in perturbation theory and 
in our scheme for small initial couplings: for $d>2$, the conditions $g_{1,\Lambda}=0$ and 
$g_{2,\Lambda}$ small make the system critical and the flow 
is driven towards the Gaussian fixed point. However, for large enough $g_{2,\Lambda}$ the flow always
becomes singular, even when $g_{1,\Lambda}=0$. We show in Fig.\ref{fig: DPcond} the basin of attraction 
of the Gaussian fixed point in this case as a function 
of $g_{2,\Lambda}$ obtained within perturbation theory. The result is similar to what is found 
in the Kardar-Parisi-Zhang equation \cite{KPZ}\footnote{In the KPZ problem,
the RG flow goes towards the Gaussian fixed point at small initial coupling for $d>2$ \cite{WieCan_SFP}
(the interface is flat), while at large coupling it goes towards a strong coupling fixed point 
(the interface is rough) \cite{Can10,Can11} and for  $d<2$ the interface is always rough.
Two is thus the critical dimension {\it at small coupling} as in PCPD.}.
Note that within perturbation theory, criticality is reached when $g_{1,\Lambda}= 0$.
It is no longer clear in our approach that this condition is necessary for large initial $g_{2,\Lambda}$ and $d>2$
because the finite scale singularity emerges for generic  $g_{1,\Lambda}$.
Our results suggest that there could exist a nontrivial critical behavior at large initial $g_{2,\Lambda}$
even above one dimension that could be in the DP (or possibly DP') universality class. 

To conclude, we believe our results represent a breakthrough for the understanding
of the critical behavior of PCPD. 
Even though we do not  have yet a complete solution of the problem, we have unlocked
an heretofore blocked situation and offered new lines of further research.
The recourse to functional nonperturbative renormalization was essential because it allows us
%, contrary to perturbative methods, 
to address questions such as the generation of linear terms. 
Going beyond our semi-functional approximation remains
the main challenge that should lead, once
controlled numerically, to satisfactory results. From a physical point of view, 
a final answer to the question of the phase diagram and its accessible fixed points in all dimensions
remains of course the main goal. But understanding  the 
meaning of the scale $k_{\rm c}$ is also challenging. It could be related to the existence of relevant ``elementary excitations"
made of pairs of particles that would involve an intrinsic scale. Disentangling the roles of the particles 
and of the pairs has already been studied by effectively taking them into account through the introduction of
 another species $B$ and the reactions:  $2 A\to B$, $B\to A$, $B\to 2B$, $2B\to B$, $B\to \emptyset$. Such two-species PCPD models, 
 which are believed to exhibit the same critical behavior as the one species case studied here \cite{PCPD_2SPECIES}, 
 could be the starting point of more complicated NPRG approaches (involving four fields, but possibly
 deprived of singularities in the flow). These endeavors are left 
 for future work.
 
 \acknowledgments
 
 We thank Matthieu Tissier and Gilles Tarjus for illuminating discussions.
 
%%%%%%%%%%%%%%%%%%%%%%%%%%%%%%%%%%%%%%%%%%%%%%%%%%%%%%%%%%%%%%%%%%%%%%%%%%%%%%%%%%%%%%%%%%%%%%%%%%%%%%%%%%%%%%%%%%%%%%%%%%%%%%%%%%%%%%%%%%%%%%

\end{document}